\documentclass[prb,floats,aps,amssymb,preprint,epsf,epsfig,rotate,superscriptaddress]{revtex4}
\usepackage{bm}
\usepackage{graphicx}
\usepackage{xcolor}

\begin{document}

\title{Semilocal approximations to the Kohn-Sham exchange potential as applied to a metal surface}

\author{C. M. Horowitz}
\affiliation{Instituto de Investigaciones Fisicoqu\'imicas Te\'oricas 
y Aplicadas, (INIFTA), UNLP, CCT La Plata-CONICET, Sucursal 4, Casilla de Correo 16, (1900)
La Plata. Argentina}
\author{C. R. Proetto}
\affiliation{Centro At\'omico Bariloche and Instituto Balseiro, 8400, S. C. de Bariloche, R\'{i}o Negro, Argentina}
\author{J. M. Pitarke}
\affiliation{CIC nanoGUNE BRTA, Tolosa Hiribidea 76, E-20018 Donostia, Basque Country, Spain}
\affiliation{Fisika Saila, Centro F\'isica Materiales CSIC-UPV/EHU, and DIPC, 644 Posta Kutxatila, E-48080 Bilbo, Basque Country, Spain}

\date\today

\begin{abstract}
    Several semilocal exchange potentials usually employed in the framework of density-functional theory (DFT) are tested and compared with their exact counterpart, the exchange Optimized Effective Potential (OEP),
    as applied to the jellium-slab model of a metal-vacuum interface. 
    Driven by their explicit dependence on the ground-state density, its gradient, and its kinetic-energy density, the three analyzed semilocal exchange potentials approach
    their respective asymptotic limits faster than in the case of the OEP, all of them having an asymptotic scaling of the form $-\alpha\,e^2/z + V_{\infty}$,
    with $\alpha < 1$. Here we provide the leading analytic asymptotics of the three model potentials under study, and we find that none of them exhibits the exact OEP slab asymptotics $-\;e^2/z$. While the so-called Becke-Roussel potential's leading asymptote is close to its exact OEP counterpart, the other two model potentials under study approach a material-dependent positive constant value far into the vacuum, resulting in considerably overestimated ionization potentials.
\end{abstract}

\maketitle

\section{Introduction}
With the introduction in 1976 of the Optimized Effective Potential (OEP) method of Talman and Shadwick,\cite{TS76} the Kohn-Sham (KS) exact exchange potential $V_x(\mathbf r)$ of Density-Functional Theory (DFT) can nowadays be calculated numerically for an arbitrary many-electron system.
The method is based on the fact that the exact KS exchange energy functional of DFT is known in terms of the KS orbitals, thus becoming an {\it implicit} energy functional of the ground-state electron density. This non-explicit dependence has the consequence that $V_x(\mathbf r)$ must be found by solving a complicated integro-differential equation.\cite{ED11}
The exchange-only ($x$-only) OEP formulation of Talman and Shadwick was later generalized to include correlation; see, for instance, the reviews in Refs.~[\onlinecite{Grabo00}] and [\onlinecite{KK08}]. 
The OEP method was originally implemented in real space to study spherical systems like 
atoms.\cite{TS76} More recently, the method was implemented for periodic solids, by using 
plane waves,\cite{SMVG97,SMMVG99} and for molecules using Gaussian basis sets.\cite{G99,IHB99}
A possible solution to the numerical instabilities that are present when using a Gaussian 
basis set has been recently proposed and successfully tested.\cite{TG21}

The computational cost of the $x$-only OEP method motivates,\cite{note10} however, the search for simpler model exchange potentials, beyond the widely used Local-Density Approximation (LDA), 
but still sharing some features of the exact $V_x(\mathbf r)$, as for example the correct $-e^2/r$ asymptotics for finite systems. Here, we consider a jellium slab and investigate the performance of three semilocal model 
exchange potentials, \cite{BR89,BJ06,RPP10} whose asymptotics we compare to those of the corresponding $x$-only OEP results. In all cases, full self-consistent convergence has been numerically achieved. The three semilocal exchange model potentials under study depend not only on the electron density but also on its gradient and 
kinetic-energy density. Indeed, this 
partial non-locality brings some important features of the exact KS exchange potential: (i) the correct $- \; e^2/r$ asymptotics (for finite systems) -in the case of two model potentials-\cite{BR89,RPP10} and (ii) a reasonable prediction, with an accuracy of about 30\,{\%},\cite{TB09} of band gaps in extended solid systems -in the case of a slightly modified version of one of the model potentials under study (usually denoted as MBJ)-.

The full non-locality of the $x$-only OEP exchange potential has also been explored to yield the correct
asymptotics of model and real solid films, which are known to be of the form $-e^2/z$ both in the case of jellium
slabs\cite{HPR06} and in the case of graphene and Si(111) films.\cite{Ye15,Engel14a,Engel14b,Engel18a,Engel18b,Engel18c} 
In the latter case, the Krieger-Li-Iafrate (KLI)
approximation\cite{KLI92} was implemented within the 
$x$-only
OEP scheme, as a way of lightening the computational cost of full {\it ab-initio} OEP calculations. The capability of simplified OEP schemes for the calculation of semiconductor work functions was also explored with the use of one of the semilocal exchange model potentials analyzed here. Seventeen semiconductors were considered, and accurate results were obtained -comparable to those obtained at the level of the more sophisticated GW approximation- with a computational cost at the level of LDA/GGA calculations.\cite{Ye16} These results were, however, debated recently in Ref. [\onlinecite{RMB21}], an issue that will be part of our discussion below.

The present work is organized as follows: in Sec. II, we give a short account of the main features of the OEP
exchange potential; in Sec. III, we present our results for three semilocal exchange model potentials, as applied to jellium slabs; and Sec. IV is devoted to the Conclusions. In the Appendix, we explain details of our analytical derivations leading to the rigorous jellium-slab asymptotics of the three semilocal model exchange potentials under study.

\section{Exact Kohn-Sham exchange potential at jellium slabs}
Our calculations are restricted to the jellium-slab model of a metal surface,
where the discrete character of the positive ions inside the metal
is replaced by a uniform distribution of positive charge (the jellium), expressed as follows:
\begin{equation}
    n_+(z) = \bar{n} \;\theta(-z) \; \theta(z + d) \; . 
\end{equation}
Here, $\bar{n}$ is a constant with the dimensions of a three-dimensional (3D) density that through the overall neutrality condition fixes the global electron density, and $d$ is the slab width. The jellium-slab model of a metal surface, with vacuum-metal interfaces at $z=-d$ and $z=0$, is defined by just these two external parameters: $\bar n$ and $d$. Taking the limit $d \rightarrow \infty$, the model reduces to the semi-infinite jellium model of a metal surface introduced by Lang and Kohn in their seminal work on DFT as applied to extended solid systems.\cite{LK70} 

The jellium-slab model is invariant under translations in
the $x$-$y$ plane, so the KS eigenfunctions can be factorized as follows:\cite{Note1}
\begin{equation}
 \varphi_{i,{\bf k}}^{\sigma}({\bf r})=\frac{e^{i{\bf k\cdot \bm{\rho} }}}{\sqrt{A}}\,
 \xi_{i}^{\sigma}(z),
 \label{KSfunctions}
\end{equation}
where ${\bm{\rho} }$ and ${\bf k}$ are the in-plane coordinate and
wavevector, respectively, and $A$ represents a normalization area.
$\xi _{i}^{\sigma}(z)$ are the normalized spin-dependent
eigenfunctions of electrons in slab discrete levels (SDL's) $i$ $(i=1,2,...)$
with energies $\varepsilon_{i}^{\sigma}.$ They are the solutions of the effective one-dimensional
KS equation (we use atomic units throughout)
\begin{equation}
\widehat{h}_{\text{KS}}^{\sigma}(z) \; \xi _{i}^{\sigma}(z)\equiv\left[ -\frac{1}{2}
\frac{\partial ^{2}}{\partial z^{2}}+V_{\text{KS}}^{\sigma}\left( z\right)\right] 
\xi _{i}^{\sigma}(z) = \varepsilon _{i}^{\sigma} \; \xi _{i}(z) \, .
\label{KSequations}
\end{equation}

In the $x$-only scenario considered here, the KS potential $V_{\text{KS}}^{\sigma}(z)$ entering Eq.~(\ref{KSequations}) is the sum of
two distinct contributions: 
\begin{equation}
 V_{\text{KS}}^{\sigma}(z) = \overline{V}_{\text{H}}(z)+V_{x,\sigma}(z) \; ,
\label{KSpotential}
\end{equation}
where $\overline{V}_{\text{H}}(z)$ is the {\it effective} electrostatic Hartree potential,\cite{Note4}
\begin{equation}
\overline{V}_{\text{H}}(z) := V_{\text{ext}}(z) + V_{\text{H}}(z)=-2\pi \int_{-\infty }^{\infty }dz^{\prime }\left|
z-z^{\prime }\right| \left[ n(z^{\prime })-n_{+}(z^{\prime })\right] ,
\label{hartree}
\end{equation}
and $V_{x,\sigma}(z)$ is the KS exchange potential, which in the OEP framework is obtained in the following way:
\begin{equation}
V_{x,\sigma}^{\text{OEP}}(z) =  V_{x,\sigma}^{\text{Slater}}(z) + V_{x,\sigma}^{\Delta}(z) + V_{x,\sigma}^{\text{Shift}}(z) \; .
 \label{VxOEP}
\end{equation}
Explicit expressions for $V_{x,\sigma}^{\text{Slater}}(z)$, $V_{x,\sigma}^{\Delta}(z)$, and $V_{x,\sigma}^{\text{Shift}}(z)$ for a slab
geometry can be found elsewhere.\cite{HPR06} In the widespread KLI approximation,\cite{KLI92} $V_{x,\sigma}^{\text{Shift}}(z)$ is neglected, so
$V_{x,\sigma}^{\text{OEP}}(z)$ reduces to $V_{x,\sigma}^{\text{KLI}}(z) := V_{x,\sigma}^{\text{Slater}}(z) + V_{x,\sigma}^{\Delta}(z)$.

The electron density $n(z)$ is obtained as follows:
\begin{equation}
n(z) = n_{\uparrow}(z) + n_{\downarrow}(z),
\end{equation}
where
\begin{equation}
n_{\sigma}(z)= \frac{1}{4\pi }\sum_{i=1}^{M_{\sigma}} \left( k_{F}^{i,\sigma} \right)^{2}
\left| \xi_{i}^{\sigma}(z)\right| ^{2}.  \label{density}
\end{equation}
Here,
$M_{\sigma}$ is the spin-dependent highest occupied slab discrete level (HOSDL), $k_{F}^{i,\sigma} = \sqrt{2(\mu -\varepsilon _{i}^{\sigma})}$, 
and $\mu$ is the chemical potential determined from the overall charge-neutrality condition
\begin{equation}
\int_{-\infty}^{\infty} [n(z)-n_+(z)] dz = 0.
 \label{neutrality}
\end{equation}
For the scope of the present work, two important exact features of the spin-compensated KS exchange potential $V_{x}^{\text{OEP}}(z) := V_{x,\uparrow}^{\text{OEP}}(z)= V_{x,\downarrow}^{\text{OEP}}(z)$, 
resulting from the self-consistent solution of 
Eqs.~(\ref{KSequations})-(\ref{neutrality}), are the following: (i) the bulk value $V_{x}^{\text{OEP}}(\text{bulk}) = - \; k_F /\pi = - \; (9/4\pi^2)^{1/3} (r_s)^{-1} $,\cite{Note2} and (ii) the asymptotic scaling $V_{x}^{\text{OEP}}(z/d \gg 1) \rightarrow - \; 1 / z$.\cite{Note5} From now on, the absence of the spin index $\sigma$ in any symbol will mean that the corresponding magnitude refers to a spin-compensated jellium slab. 

\begin{figure}
 \includegraphics[width=12cm]{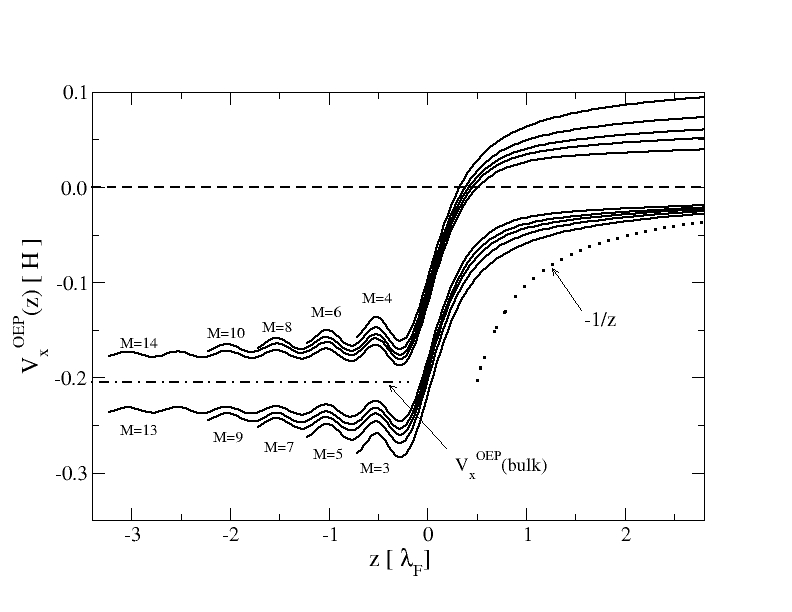}
 \caption{Self-consistent OEP calculations of the KS exact exchange potential of
 Eq.~(\ref{VxOEP}) for $r_s=3$ and jellium slabs with a number $M$ of occupied SDLs going from $M=3$ ($d=1.53  \; \lambda_F$) to $M=14$ ($d = 6.55  \; \lambda_F$). The slab width $d$ has been chosen carefully in such a way that (i) the filling factor $\eta_M\sim 1^-$ ($k_F^M\sim 1/d$) when $M$ is odd and (ii) the filling factor
 $\eta_M\sim 0^+$ ($k_F^M\to 0$) when $M$ is even. The bulk limit for $r_s = 3$ is represented by a dashed-dotted line, and the vacuum asymptotic limit $V_x^{\text{OEP}}(z/d \gg 1) \rightarrow -1/z$ is represented by a dotted curve. 
In all cases, the right metal-vacuum interface is at $z=0$.}
\label{grafico.12.jpg}
\end{figure}

\begin{figure}
 \includegraphics[width=12cm]{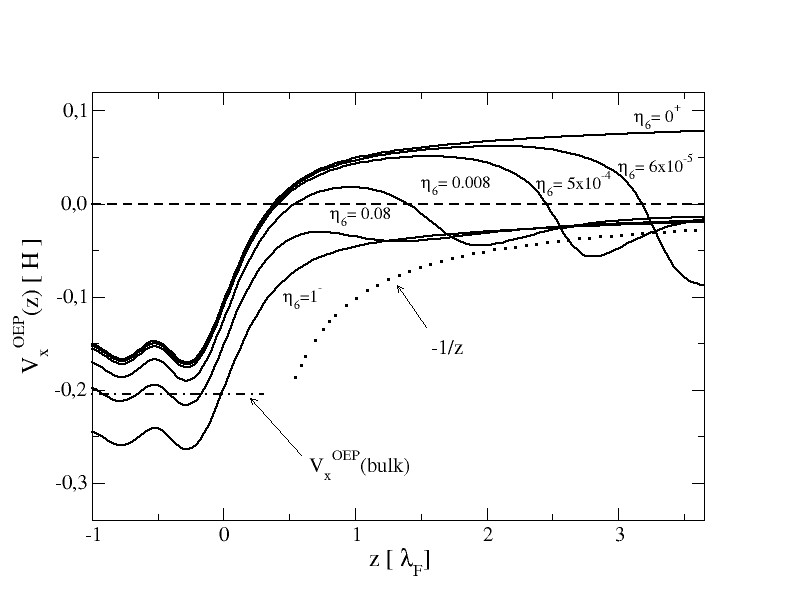}
 \caption{Self-consistent OEP calculations of the KS exact exchange potential of
 Eq.~(\ref{VxOEP}) for $r_s=3$ and jellium slabs with $M=6$ and various values of the slab thickness $d$ corresponding to filling factors that go from $\eta_6=0^+$ ($d=2.5435 \; \lambda_F$) to $\eta_6=1^-$ ($d=3.0472 \; \lambda_F$). The bulk limit for $r_s = 3 $ is represented by a dashed-dotted line, and the vacuum asymptotic limit $V_x^{\text{OEP}}(z/d \gg 1) \rightarrow -1/z$ is represented by a dotted curve. 
As in Fig. 1, $z=0$ represents the right metal-vacuum slab interface.}
\label{grafico.13.jpg}
\end{figure}

Figure~\ref{grafico.12.jpg} shows self-consistent OEP calculations of the KS exact exchange potential of Eq.~(\ref{VxOEP}) for jellium slabs with a number $M$ of occupied SDLs that goes from $M=3$ to $M=14$. The slab width $d$ has been chosen carefully in such a way that either (i) a new SDL is just about to be occupied (high filling factor $\eta_M$; $k_F^M\sim 1/d$) or (ii) a new SDL has just been occupied (low filling factor $\eta_M$; $k_F^M\to 0$), with the filling factor $\eta_M$ being defined as follows:
\begin{equation}
    \eta_M = \frac{\mu-\varepsilon_M}{\varepsilon_{M+1} - \varepsilon_M} > 0 \; .
\end{equation}
For a given $M$, $\varepsilon_M$ is the highest occupied SDL and $\varepsilon_{M+1}$ is the lowest unoccupied 
SDL, so $\eta_M$ takes values between $0^+$ (a new SDL has just been occupied) and $1^-$ (a new SDL is just about to be occupied). In Fig.~\ref{grafico.12.jpg}, a slab thickness corresponding to high filling factors ($\eta_M\sim 1^-$; $k_F^M\sim 1/d$) has been chosen when $M$ is odd, and a slab thickness corresponding to low filling factors
($\eta_M\sim 0^+$; $k_F^M\to 0$) has been chosen when $M$ is even. The result is that when a new SDL is just about to be occupied ($\eta_M\sim 1^-$; $k_F^M\sim 1/d$), the KS exact exchange potential approaches (i) the bulk limit from below as $d$ {\it increases} and (ii) the expected slab asymptotics $-1/z$ as $d$ {\it decreases}. Instead, when a new SDL has just been occupied ($\eta_M\sim 0^+$; $k_F^M\to 0$), the KS exchange potential (i) approaches the bulk limit from above as $d$ {\it increases} and (ii) {\it never} approaches
the $-1/z$ asymptotics. The reason for this is that when a new SDL has just been occupied ($\eta_M\sim 0^+$; $k_F^M\to 0$), 
the necessary condition for approaching the asymptotic regime $k_F^M\,z \gg 1$ is never reached.\cite{Note5}

Figure~\ref{grafico.13.jpg} displays self-consistent OEP calculations of the KS exact exchange potential of Eq.~(\ref{VxOEP}) for jellium slabs with $M=6$ and various values of the slab thickness $d$ corresponding to filling factors that go from $\eta_6=0^+$, in which case the slab asymptotics is never reached, to $\eta_6=1^-$, in which case the slab asymptotics ($k_F^M\,z \gg 1$) is nicely reached as $z/d \gg 1$.  

It is important to address the fact that the remarkable finite-size effects displayed by 
$V_x^{\text{OEP}}(z)$ in Figs. 1 and 2, while real, have been somehow overemphasized by our 
careful choice of the slab width $d$. Taking, for instance, Fig. 2, one observes that considerable 
finite-size effects occur only for filling factors over $0.1$, which leads us to the conclusion 
that in about 90 percent of all possible situations finite-size effects are either absent 
($\eta_6 \sim 1^-$) or very small ($1 > \eta_6 > 0.1$). In any case, these finite-size effects 
need to be carefully analyzed, as they exhibit dramatically a rigorous property of the exact 
slab exchange potential. 

At this point, it is interesting to point out that the OEP calculation of the KS exchange potential
$V_x^{\text{OEP}}(z)$ simplifies dramatically in the extreme quantum limit of one single occupied SDL 
(M=1), first introduced in Ref.~[\onlinecite{RP03}]. One can prove that in this situation 
$V_x^{{\Delta}}(z) \equiv V_x^{\text{Shift}}(z) \equiv 0$, so the Slater potential $V_x^{\text{Slater}}\equiv 2\,\varepsilon_x(z)$ remains the only 
surviving contribution in Eq.~(\ref{VxOEP}), with $\varepsilon_x(z)$ being the position-dependent exchange energy per particle.\cite{HPR06,HPP08,HCPP09} This analytical approach, within the $x$-only OEP framework, to the singly-occupied SDL regime was later generalized to lower dimensions\cite{N16} and extended to the domain of time-dependent DFT (TDDFT).\cite{N17}

\section{Semilocal approximations}

The $x$-only OEP approach to the calculation of the KS exchange potential of DFT involves the numerical study of a complicated integro-differential equation.\cite{TS76}
The difficulty is rooted in the fact that the exchange energy functional (Fock-integral) on which the OEP method is based is an {\em explicit} functional of the KS orbitals but only an {\em implicit} functional of the ground-state electron density. This is in contrast with the often-used local-density approximations, in which case one easily obtains the KS exchange (and exchange-correlation) potential from the knowledge of the electron density. It is then quite natural to find out that several functionals with 
different degrees of semilocality (GGA, meta-GGA, etc.) have been formulated over the years as a way to obtain accurate KS exchange (and exchange-correlation) model potentials without the burden of solving the OEP equations. Here, we analize {\it three}\cite{BR89,BJ06,RPP10} semilocal KS exchange functionals, which we
apply to the metal surface and compare to the KS exact exchange potential that we obtain by using the OEP method, with a particular emphasis on the long-range asymptotic behavior.

\subsection{Becke-Roussel exchange potential $V_{x,\sigma}^{\text{BR}}(\mathbf{r})$}

This approximation to the KS exchange potential is based on the use of the spherically-averaged exchange hole of the three-dimensional (3D) hydrogen atom. 
As such, it should be considered as an approximation to the Slater contribution entering 
Eq.~(\ref{VxOEP}), neglecting both $V_{x,\sigma}^{\Delta}(z)$ and 
$V_{x,\sigma}^{\text{Shift}}(z)$. It includes some features of the KS
exact exchange potential in the limits of a uniform electron system and the hydrogen atom, 
to be discussed in more detail below, and it yields the correct $-1/r$ asymptotics for finite systems. The so-called Becke-Roussel (BR) exchange potential is defined by the following set of equations:\cite{BR89}

\begin{equation}
 V_{x,\sigma}^{\text{BR}}(\mathbf{r}) = -\frac{1}{b_{\sigma}(\mathbf{r})}\left[1-e^{-{x_{\sigma}}(\mathbf{r})}-
 \frac{1}{2}{x_{\sigma}}(\mathbf{r})e^{-{x_{\sigma}}(\mathbf{r})}\right]  \; ,
 \label{BR-1}
\end{equation}

\begin{equation}
 b_{\sigma}^3(\mathbf{r}) = \frac{{x_{\sigma}^3}(\mathbf{r})
 e^{-{x_{\sigma}}(\mathbf{r})}}{8 \pi n_{\sigma}(\mathbf{r})}  \; ,
 \label{BR-2}
\end{equation}

\begin{equation}
 \frac{{x_{\sigma}}(\mathbf{r})e^{-(2/3){x_{\sigma}}(\mathbf{r})}}
 {{x_{\sigma}}(\mathbf{r})-2}= \frac{2}{3} \pi^{2/3} \frac{n_{\sigma}^{5/3}(\mathbf{r})}
 {Q_{\sigma}(\mathbf{r})} \; ,   
 \label{BR-3}
\end{equation}

\begin{equation}
 Q_{\sigma}(\mathbf{r})= \frac{1}{6} [\nabla^2 n_{\sigma}(\mathbf{r})-2 \gamma D_{\sigma}(\mathbf{r})] \; ,
 \label{BR-4}
\end{equation}

\begin{equation}
 D_{\sigma}(\mathbf{r})= t_{\sigma}(\mathbf{r})-\frac{1}{4} \frac{[\nabla n_{\sigma}(\mathbf{r})]^2}
 {n_{\sigma}(\mathbf{r})} \; , 
 \label{BR-5}
\end{equation}

\begin{equation}
 t_{\sigma}(\mathbf{r})= \sum_{i,\mathbf{k}}^{occ} |\nabla \varphi_{i,{\bf k}}^{\sigma}({\bf r})|^2 \; . 
 \label{BR-6}
\end{equation}

Here, $t_{\sigma}(\mathbf{r})$ represents (twice) the spin-dependent kinetic-energy density and $D_{\sigma}(\mathbf{r})$ is a well-known quantity that is present in the so-called electron-localization function\cite{BE90,BMG05,RCG08} and also enters the expression of the local curvature of the exchange hole.\cite{D93} 
A detailed comparison between the Slater and the BR exchange potentials in solids was
presented in Ref.~[\onlinecite{TBS15}]. A test set including semiconductors and insulators of 
various types was considered, and it was concluded that these potentials yield electronic structures 
that are very similar to each other. However, in a few cases, as in the strongly correlated system 
NiO, the fundamental band gap or magnetic properties can differ significantly.

Introducing the factorized KS orbitals of Eq.~(\ref{KSfunctions}) into Eq.~(\ref{BR-6}) and performing a two-dimensional integral over
the occupied $\mathbf{k}$'s, one obtains:
\begin{equation}
  t_{\sigma}(\mathbf{r}) = t_{\sigma}(z) = 
  \sum_{i=1}^{M_{\sigma}} \frac{(k_F^{i,\sigma})^4}{8\pi} [\xi_{i}^{\sigma}(z)]^2 +
  \sum_{i=1}^{M_{\sigma}} \frac{(k_F^{i,\sigma})^2}{4\pi} \left[\frac{d\xi_{i}^{\sigma}(z)}{dz}\right]^2 \; .
  \label{ts}
\end{equation}
Now we insert this expression into Eq.~(\ref{BR-5}), and using Eq.~(\ref{density}) one finds:
\begin{eqnarray}
 D_{\sigma}(\mathbf{r}) = D_{\sigma}(z) &=& \sum_{i=1}^{M_{\sigma}} \frac{(k_F^{i,\sigma})^4}{8\pi} [\xi_{i}^{\sigma}(z)]^2 +
 \sum_{i=1}^{M_{\sigma}} \frac{(k_F^{i,\sigma})^2}{4\pi} \left[\frac{d\xi_{i}^{\sigma}(z)}{dz}\right]^2 \nonumber  \\
 &-& 
 \sum_{i,j=1}^{M_{\sigma}} \frac{(k_F^{i,\sigma} k_F^{j,\sigma})^2}{16\pi^2} \frac{\xi_{i}^{\sigma}(z)\xi_{j}^{\sigma}(z)}{n_{\sigma}(z)} 
 \frac{d\xi_{i}^{\sigma}(z)}{dz} \frac{d\xi_{j}^{\sigma}(z)}{dz}.
 \label{Ds}
\end{eqnarray}
Finally, one obtains:
\begin{equation}
    Q_{\sigma}(\mathbf{r}) = Q_{\sigma}(z) = \frac{1}{6} \left[
    \sum_{i=1}^{M_{\sigma}} \frac{(k_F^{i,\sigma})^2}{2\pi}\left[\xi_{i}^{\sigma}(z) \frac{d^2\xi_{i}^{\sigma}(z)}{dz^2} +\frac{d\xi_{i}^{\sigma}(z)}{dz} \frac{d\xi_{i}^{\sigma}(z)}{dz} \right] - 2 \gamma D_{\sigma}(z)  \right].
    \label{Qs}
\end{equation}
Here, $\gamma$ is a dimensionless parameter to be determined below by imposing the constraint that the bulk value of $V_{x,\sigma}^{\text{BR}}(z)$ should agree with the bulk value of the Slater potential $V_{x,\sigma}^{\text{Slater}}(z)$. 
This is discussed in detail in Appendix A.
It should be noted already at this point that, in the limit $z/d \gg 1$, all the sums over occupied SDLs collapse to the HOSDL for each $M_{\sigma}$; for instance, $n_{\sigma}(z/d \gg 1) \rightarrow (k_F^{M_{\sigma}})^2 \xi_{M_{\sigma}}^{\sigma}(z/d \gg 1)^2 / (4\pi)$, with $k_F^{M_{\sigma}} = \sqrt{2(\mu-\varepsilon_{M_{\sigma}}})$. This collapse of all quantities towards the HOSDL is the key for obtaining analytically the asymptotic limit of $V_{x,\sigma}^{\text{BR}}(z)$, as we explain below.
On the other side, this assumption is not valid for the semi-infinite geometry 
($d \rightarrow \infty$ in our slab model), leading for instance to a qualitatively different 
asymptotic limit of the exact exchange potential. This has been discussed in detail recently 
by us in Ref. [\onlinecite{HPP21}].

It is also worth noting that all quantities involved in the determination of the jellium-slab BR exchange potential become {\it effective} one-dimensional magnitudes after integration over the in-plane degrees of freedom, as expected. As a consequence, the BR 
slab exchange potential itself reduces to an effective one-dimensional magnitude, as follows:
\begin{equation}
     V_{x,\sigma}^{\text{BR}}(z) = - \sqrt{6} \left[\frac{Q_{\sigma}(z)}{x_{\sigma}(z)-2} \times \frac{1}{x_{\sigma}(z)n_{\sigma}(z)}\right]^{1/2}
     \left[ 1-e^{-{x_{\sigma}}(z)}-\frac{1}{2}{x_{\sigma}}(z)e^{-{x_{\sigma}}(z)} \right].
     \label{BR1D}
\end{equation}

\begin{figure}
 \includegraphics[width=12cm]{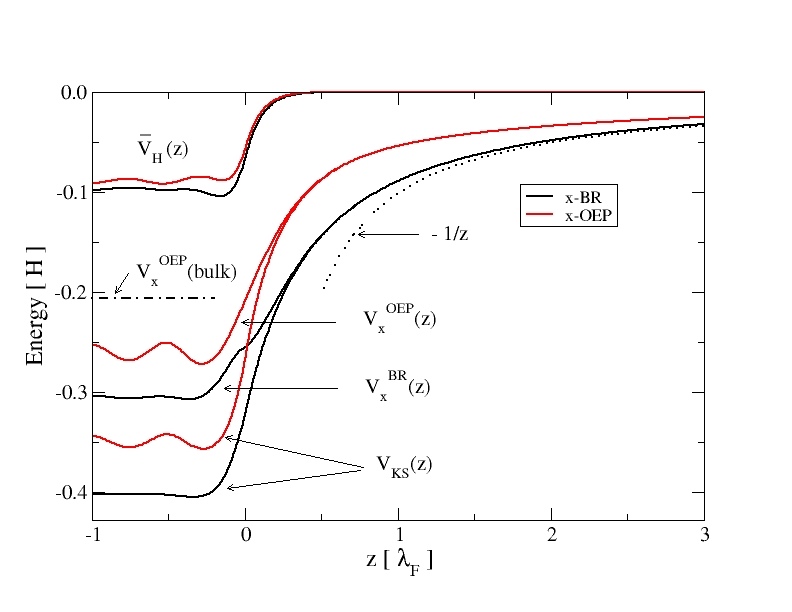}
 \caption{Self-consistent OEP (red) and BR (black) evaluations of the KS exchange potential, for $r_s=3$ and $d = 2 \; \lambda_F$. For these values of $r_s$ and $d$, $M=4$ and $\eta_4 \sim 0.8$. The corresponding $\overline{V}_{\text{H}}(z)$ and $V_{\text{KS}}(z)$ potentials are also represented for comparison. The OEP bulk limit
 ($V_x^{\text{OEP}}(\text{bulk}) \simeq -\;0.204$) is represented by a dashed-dotted line. The vacuum asymptotic limit $V_x^{\text{OEP}}(z/d \gg 1) \rightarrow -1/z$ is represented by a dotted curve.}
\label{Fig:g1}
\end{figure}

In Fig.~\ref{Fig:g1}, we display a comparison between self-consistent calculations of $V_x^{\text{BR}}(z)$ and $V_x^{\text{OEP}}(z)$, together with the corresponding $\overline{V}_{\text{H}}(z)$ and $V_{\text{KS}}(z)$ potentials. While
$V_x^{\text{OEP}}(z)$ is obtained from the self-consistent solution of Eqs.~(\ref{KSequations})-(\ref{VxOEP}), $V_{x}^{\text{BR}}(z)$ is obtained, instead, from the self-consistent solution of Eqs.~(\ref{KSequations})-(\ref{hartree}) by introducing into Eq.~(\ref{KSpotential}) the Becke-Roussel exchange potential $V_{x}^{\text{BR}}(z)$ of Eq.~(\ref{BR1D}) instead
of the actual KS exchange potential of Eq.~(\ref{VxOEP}). The effective electrostatic Hartree potential $\overline{V}_{\text{H}}(z)$ is found to be reasonably well approximated in the present BR model; but the entire KS exchange potential is considerably 
deeper in this model, particularly in the bulk side of the surface. This substantial bulk discrepancy is simply due to the fact that the BR potential is an approximation to the Slater potential $V_{x}^{\text{Slater}}(z)$ (see the Appendix), which is well known to be too negative in the bulk by a factor of 3/2. Figure~\ref{Fig:g1} also shows that
$V_{x}^{\text{BR}}(z)$ is less affected by Friedel-like oscillations and
exhibits a kind of a kink at $z=0$ (right at the metal-vacuum interface). As for the vacuum asymptotics, $V_{x}^{\text{BR}}(z)$ displays a scaling of the form 
$-\alpha^{\text{BR}}/z$, as in the case of the exact $V_{x}^{\text{OEP}}(z/d \gg 1)\to-1/z$, but now with 
$\alpha^{\text{BR}}<1$, as shown below.

A comparison between $V_{x}^{\text{BR}}(z)$ and $V_{x}^{\text{Slater}}(z)$ is provided in Fig.~\ref{Fig:g1b}.
Considering that the interface is at $z=0$, this figure shows that both potentials remain close inside the slab, which is a consequence of the fulfillment of the bulk constraint  
$V_{x}^{\text{BR}}(\text{bulk}) \simeq V_{x}^{\text{Slater}}(\text{bulk}) = -0.306$, for $r_s = 3$; however, they differ appreciably in the near-interface vacuum region, with the exact ($x$-OEP generated) Slater exchange potential
being more localized than its Becke-Roussel counterpart. We attribute this feature to the 
only-partial non-locality of
$V_{x}^{\text{BR}}(z)$, which results in a faster and closer approach to the 
correct $-1/z$ asymptotics, as opposed to the exact $V_{x}^{\text{Slater}}(z)$, whose full 
non-locality results in a much slower approach to the universal $-1/z$ asymptotics. Indeed, the Slater potential $V_{x}^{\text{Slater}}(z)$ is built from the exact exchange-hole, while $V_{x}^{\text{BR}}(z)$ is constructed from the spherically averaged exchange-hole of the hydrogen atom, which cannot possibly account for the fact that on the vacuum side of the surface the actual exchange hole is left behind and far from the electron itself.\cite{CP09,RRP03} The result is that the screening capability of the BR exchange hole is asymptotically too large and the absolute value of $V_{x}^{\text{BR}}(z)$ is, therefore, asymptotically smaller than 
$V_{x}^{\text{Slater}}(z)$ (see the inset of Fig.~\ref{Fig:g1b}), which explains the fact that the coefficient $\alpha^{\text {BR}}$ is 
ultimately smaller than unity. 

In order to obtain a rigorous analytical expression for the vacuum asymptotics of the BR model potential, we first look at the asymptotic behavior of the electron density $n_{\sigma}(z)$, which we obtain from Eq.~(\ref{KSequations}). Far into the vacuum, 
Eq.~(\ref{KSequations}) can be written as $(-\;{\partial}^2 / {\partial z}^2 - 2 \varepsilon_i^{\sigma})\xi_i^{\sigma}(z) = 0$, whose solution is $\xi_i^{\sigma}(z/d \gg 1) \rightarrow \sqrt{A_i^{\sigma}} \; e^{-z\beta_i^{\sigma}}$, with 
$\beta_i^{\sigma}=\sqrt{-2\varepsilon_i^{\sigma}}$ and $A_i^{\sigma}$ being a normalization constant along the $z$-direction.
One finds $n_{\sigma}(z/d \gg 1) \rightarrow [k_F^{M_{\sigma}} \xi_{M_{\sigma}}(z/d \gg 1)]^2/4\pi \sim A_i^{\sigma} (k_F^{M_{\sigma}})^2 e^{-2z\beta_{M_{\sigma}}}/4\pi$, and one can then derive, as indicated in the Appendix, asymptotic expansions for $t_{\sigma}(z)$,
$D_{\sigma}(z)$, $Q_{\sigma}(z)$, and $b_{\sigma}(z)$ [see Eq.~(\ref{bsa})], and finally:

\begin{figure}
 \includegraphics[width=12cm]{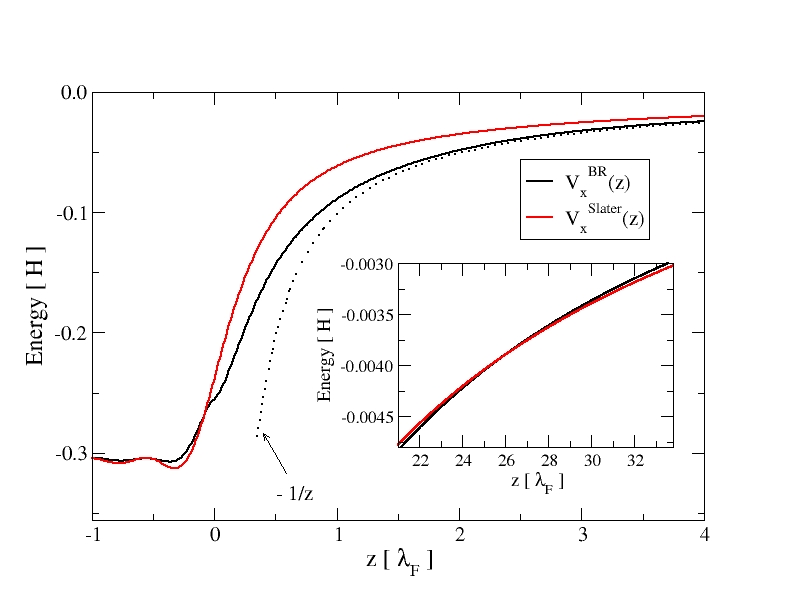}
 \caption{Self-consistent $V_x^{\text{BR}}(z)$ (black) and $V_x^{\text{Slater}}(z)$ (red) potentials, with the latter being extracted from
 $V_x^{\text{OEP}}(z) = V_x^{\text{Slater}}(z) + V_x^{\Delta}(z) + V_x^{\text{Shift}}(z)$. The inset corresponds to the far asymptotic region showing
 the crossing of $V_x^{\text{BR}}(z)$ and $V_x^{\text{Slater}}(z)$.
 Slab width $d = 2 \; \lambda_F$ and $r_s = 3$. The dotted line corresponds to the universal limit $-1/z$.}
 \label{Fig:g1b}
\end{figure}

\begin{eqnarray}
 V_{x,\sigma}^{\text{BR}}(z/d \gg 1) &=& 
 -\frac{1}{z} \left[1-\frac{\gamma (k_F^{M_{\sigma}})^2} {4 \beta_{M_{\sigma}}^2}\right]^{1/2}
 \left(1+\frac{1}{2 z \beta_{M_{\sigma}}}\right) \nonumber \; \\
 &=& - \frac{\alpha^{\text{BR}}}{z} + \mathcal{O}(z^{-2}) \; ,
 \label{BR-bulk}
\end{eqnarray}
with
\begin{equation}
\alpha^{\text{BR}} = \left[1-\frac{\gamma (k_F^{M_{\sigma}})^2} {4 \beta_{M_{\sigma}}^2}\right]^{1/2},
\label{alphabr}
\end{equation}
where $(k_F^{M_{\sigma}}) = \sqrt{2(\mu - \varepsilon_{M_{\sigma}}^{\sigma})}$ and $\beta_{M_{\sigma}} =\sqrt{V_\infty^{\text{BR}}- 2 \varepsilon_{M_{\sigma}}^{\sigma}}$, 
with $V_\infty^{\text{BR}} = 0$. Hence, we find a material-dependent scaling coefficient $\alpha^{\text{BR}}<1$, which is in contrast with the exact universal asymptotics of the form $-1/z$. These slight differences in the asymptotics are shown in Fig.~\ref{Fig:0}, for a slab thickness chosen in such a way that $M=3$ and the HOSDL is far enough from being {\it just} occupied, 
as a way of maximizing the difference between the displayed curves. Figure~\ref{Fig:0} also shows that far enough from the surface into the vacuum the BR model potential
$V_{x}^{\text{BR}}(z)$ is very well described by a potential of the form $-\alpha^{\text {BR}}/z$ with the coefficient $\alpha^{\text {BR}}$ given by Eq.~(\ref{alphabr}). 
The slab analytical asymptotics of Eqs.~(\ref{BR-bulk})-(\ref{alphabr}) represent one of the main results of the present work.

\begin{figure}
 \includegraphics[width=12cm]{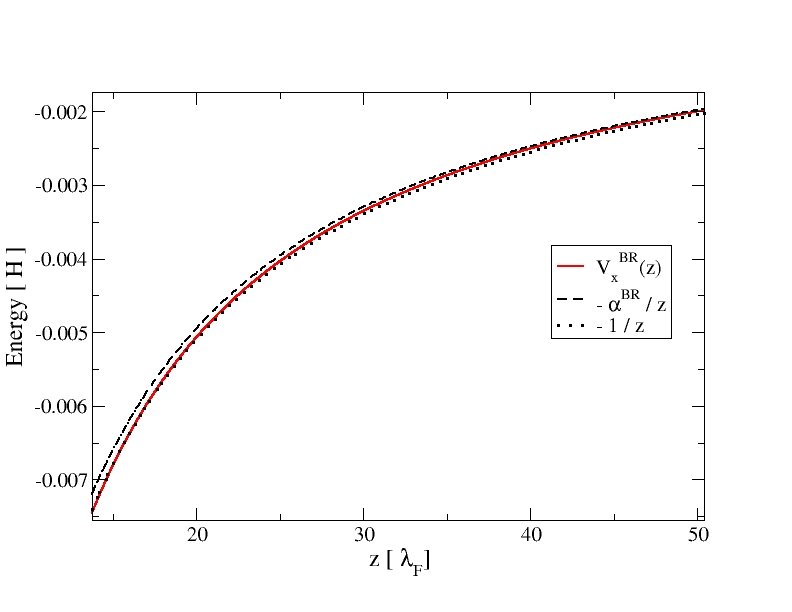}
 \caption{Numerical check of Eq.~(\ref{BR-bulk}). Self-consistent $V_x^{\text{BR}}(z)$ (red full line), $-1/z$ (dotted line), and asymptotic expression $V_x^{\text{BR}}(z/d \gg 1)$ (dashed line) as given by Eq.~(\ref{BR-bulk}) with $\alpha^{\text{BR}} = 0.9724$. Slab width $d = 1.486 \; \lambda_F$, $r_s = 3$, for which $M = 3$.}
 \label{Fig:0}
\end{figure}

As for the kink in the BR exchange potential that is visible in Figs.~3 and \ref{Fig:g1b}, we note that it arises from the factor inside the square root in Eq.~(\ref{BR1D}).
As $Q_{\sigma}(\text{bulk}) < 0$ and $Q_{\sigma}(z/d \gg 1) > 0$, the quantity $Q_{\sigma}(z)$ passes through zero at some intermediate $z=z_0$. When this happens, $x_{\sigma}(z_0) = 2$, as to keep finite $V_x^{\text{BR}}(z_0)$ of Eq.~(\ref{BR1D}). According to Fig.~\ref{Fig:x} (of the Appendix), $Q_{\sigma}(z_0) = 0$ at $z=z_0 \simeq 0$, right at the metal-vacuum interface. Hence, assuming that the ratio $Q_{\sigma}(z)/[x_{\sigma}(z)-2]$ remains finite and compensated at
$z \sim z_0$, the behavior of the BR model potential right at the interface depends on the product $x_{\sigma}(z)n_{\sigma}(z)$ at
$z\sim z_0 \sim 0$. Figure~\ref{Fig:x} (of the Appendix) shows that $x_{\sigma}(z)$ increases with $z$ while $n_{\sigma}(z)$ decays with $z$ into the vacuum: we have checked, however, that the product of these two quantities has a local maximum at $z \sim z_0 \sim 0$, explaining 
the presence of the kink in $V_x^{\text{BR}}(z)$, which should, therefore, be considered as an artefact coming from the use of the exchange hole of the hydrogen atom as a reference system.

Finally, we note (see Figs.~\ref{Fig:g1} and \ref{Fig:g1b}) that $V_x^{\text{BR}}(z)$ approaches the vacuum asymptotics considerably faster than $V_x^{\text{OEP}}(z)$. 
As already discussed above, this feature is connected with the partial locality of the BR model potential, which depends explicitly on the electron density, its gradient, and its
 kinetic-energy density. We will see below that this feature is inherited by the other two semilocal exchange potentials under study, which are both generated on the basis of the BR model.

\subsection{Becke-Johnson exchange potential $V_{x,\sigma}^{\text{BJ}}(z)$}

\begin{figure}
 \includegraphics[width=12cm]{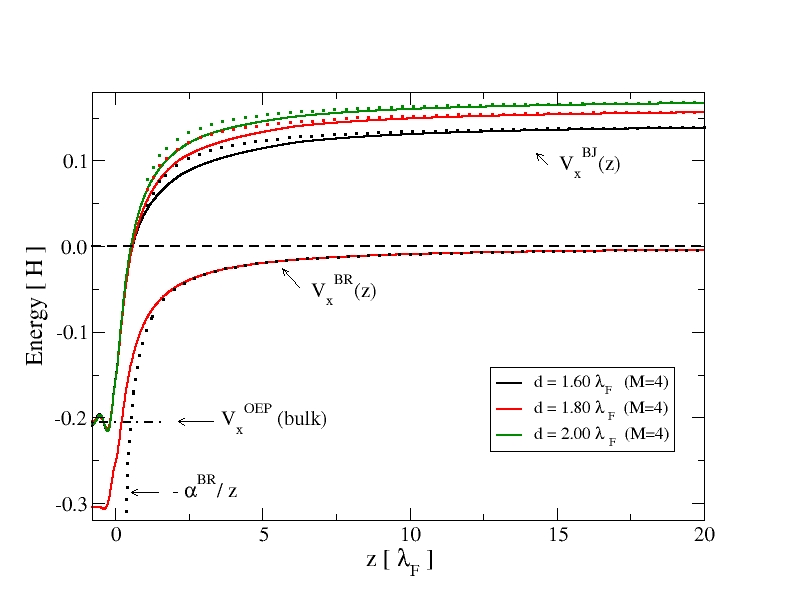}
 \caption{Upper curves: BJ exchange potential for three values of the slab 
width $d$ (full lines), and their corresponding
 asymptotics, Eq.~(\ref{BJa}) (dotted curves). Lower curves: BR exchange potential, 
for $d = 1.80 \; \lambda_F$ (full red curve), and its corresponding asymptotics (dashed curve). 
$r_s=3$, and in all cases $M = 4$ and the metal-vacuum interface is located at $z=0$.}
 \label{Fig:BJ}
\end{figure}
 
\begin{figure} 
 \includegraphics[width=12cm]{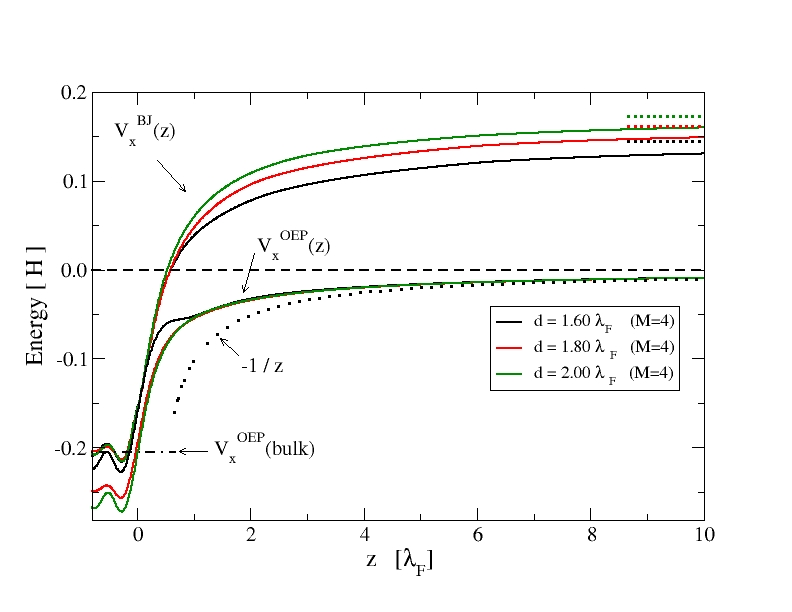}
 \caption{Becke-Johnson and $x$-only OEP exchange potentials, for different slab widths $d$, and $r_s = 3$. The upper right dotted segments are the $V_{\infty}^{\text{BJ}}$ corresponding limits, as given by Eq.~(\ref{VinftyBJ}).
  The dotted line corresponds to the universal limit $-1/z$. In all cases, $M = 4$ and the metal-vacuum interface is at $z=0$.}
 \label{Fig:BJ.OEP}
\end{figure}

The main drawback of the $V_{x,\sigma}^{\text{BR}}(z)$ model potential, as applied to a jellium slab, is the fact that it fits, in the bulk, the Slater potential, which is wrong by a factor of 3/2. This flaw was mitigated with the introduction of the Becke-Johnson exchange potential,\cite{BJ06} which in 
our case of translational invariance in two directions reads as follows:\cite{Note3}
\begin{equation}
 V_{x,\sigma}^{\text{BJ}}(z) = V_{x,\sigma}^{\text{BR}}(z) + C \left 
 [\frac{t_{\sigma}(z)}{n_{\sigma}(z)} \right]^{1/2} \; .
 \label{BJ}
\end{equation}
By choosing  $C=[{5}/({12 \pi^2})]^{1/2}$, this model exchange potential reproduces
(for $\gamma=0.8$, see Appendix) the uniform-electron-gas limit:
$V_{x,\sigma}^{\text{BJ}}(\text{bulk}) \simeq V_{x,\sigma}^{\text{OEP}}(\text{bulk}) = - [{9}/({4\pi^2})]^{1/3}/r_s$.
As the BR model potential $V_{x,\sigma}^{\text{BR}}(z)$ simply approximates $V_{x,\sigma}^{\text{Slater}}(z)$, the correction term in Eq.~(\ref{BJ}) can be
interpreted as an approximation to the contribution
$V_{x,\sigma}^{\Delta}(z) + V_{x,\sigma}^{\text{Shift}}(z)$
entering Eq.~(\ref{VxOEP}). From Eq.~(\ref{A6}), we find 
$t_{\sigma}(z/d \gg 1) / n_{\sigma}(z/d \gg 1) \rightarrow \bar{\beta}_{M_{\sigma}} + (k_F^{M_{\sigma}})^2/2$; hence, we obtain:
\begin{eqnarray}
  V_{x,\sigma}^{\text{BJ}}(z/d \gg 1) &\rightarrow&
  -\frac{1}{z} \left[1-\frac{\gamma (k_F^{M_{\sigma}})^2}
 {4 \bar{\beta}_{M_{\sigma}}^2}\right]^{1/2} +
 C \left[\bar{\beta}_{M_{\sigma}} + (k_F^{M_{\sigma}})^2/2 \right]^{1/2} \nonumber \; . \\
 &=:& -\frac{\alpha^{\text{BR}}}{z} + V_{\infty}^{\text{BJ}} \; ,
 \label{BJa}
\end{eqnarray}
where $\bar{\beta}_{M_{\sigma}} = \sqrt{2(V_{\infty}^{\text{BJ}} -\varepsilon_{M_{\sigma}})}$. Solving for $V_{\infty}^{\text{BJ}}$, we find:
\begin{equation}
    V_{\infty}^{\text{BJ}} = C^2 \left[ 1 + \sqrt{1+\frac{1}{C^2}\left(-2\varepsilon_{M_{\sigma}} + \frac{(k_F^{M_{\sigma}})^2}{2}
    \right)} \right] \; .
    \label{VinftyBJ}
\end{equation}
Far into the vacuum, the BJ slab exchange potential approaches a positive, material-dependent 
constant
$V_{\infty}^{\text{BJ}}$. Equation~(\ref{VinftyBJ}) is similar to the expression obtained in the 
case of finite systems,\cite{RMB21} the only difference being the presence of the extra term 
$(k_F^{M_{\sigma}})^2/2$ inside the square root in our case, which is finite
along the direction $z$ (this localization being the source of the
$-2\,\varepsilon_{M_{\sigma}}$ contribution),
but extended in the $x-y$ plane.


We display in Figs.~\ref{Fig:BJ}  and \ref{Fig:BJ.OEP} the Becke-Johnson model potential $V_{x}^{\text{BJ}}(z)$, for several slab widths, together with
$V_{x}^{\text{BR}}(z)$ (in Fig.~\ref{Fig:BJ}),
$V_{x}^{\text{OEP}}(z)$ (in Fig.~\ref{Fig:BJ.OEP}), and the corresponding vacuum asymptotics (dotted lines). The BJ model potential reproduces the correct 
slab bulk limit; but it fails badly to describe the actual exchange potential on the vacuum side of the surface.\cite{note9} The 
slab ionization potential or work function $W$, defined as
\begin{equation}
    W^i(d) = V_{\infty}^i(d) - \mu \; ,
\end{equation}
gives us a complementary piece of information, with $i=$ OEP, BR, BJ, and RPP (R\"as\"anen-Pittalis-Proetto, see next sub-section). Since 
$V_{\infty}^{\text{OEP}} = V_{\infty}^{\text{BR}} = 0$, then $W^{\text{OEP}}$ and $W^{\text{BR}}$ are both equal to $-\mu$, although the respective chemical
potentials are of course different. Proceeding in this way, we obtain $W^{\text{BJ}}(d = 1.6 \; \lambda_F) \approx 0.2377$, $W^{\text{BJ}}(d = 1.8 \; \lambda_F) \approx 0.2606$, and
$W^{\text{BJ}}(d = 2.0 \; \lambda_F) \approx 0.2632$, while $W^{\text{OEP}}(d = 1.6 \; \lambda_F) \approx 0.1053$,
$W^{\text{OEP}}(d = 1.8 \; \lambda_F) \approx 0.1410$, and $W^{\text{OEP}}(d = 2.0 \; \lambda_F) \approx 0.1450$.
Taking the $x$-only OEP work function as a benchmark, the BJ approximation leads to a severe overestimation of the corresponding work
function, by a factor close to two. This is a direct consequence of the incorrect asymptotic limit $V_{\infty}^{\text{BJ}} \neq 0$
displayed in Figs. 6 and \ref{Fig:BJ.OEP}.

The fact that the BJ exchange potential has a system-dependent limiting value 
far outside finite systems like atoms and molecules has some anomalous consequences 
that were analyzed in detail in Refs.~[\onlinecite{AKK17a}] and [\onlinecite{AKK17b}]. 
In particular, it was found that it has a divergent behavior in the vicinity of nodal surfaces, 
which in turn poses a challenge for the convergence of numerical solutions of the corresponding 
KS equations.

In the case of extended systems like bulk solids this system-dependent constant is not relevant, 
and the BJ exchange potential performs well for a set of selected solids 
(C, Si, BN, MgO, CuO$_2$, and NiO), as far as total energies, electronic structure, 
electric-field gradients, and magnetic moments are concerned.\cite{TBBB15}
An attempt was made in Ref.~[\onlinecite{TBBB16}] to parametrize the BJ semilocal 
exchange potential for solids using empirical parameters in order to obtain better agreement with 
the exact exchange potential.

\subsection{R\"as\"anen-Pittalis-Proetto exchange potential $V_{x,{\sigma}}^{\text{RPP}}(z)$}

\begin{figure} 
 \includegraphics[width=12cm]{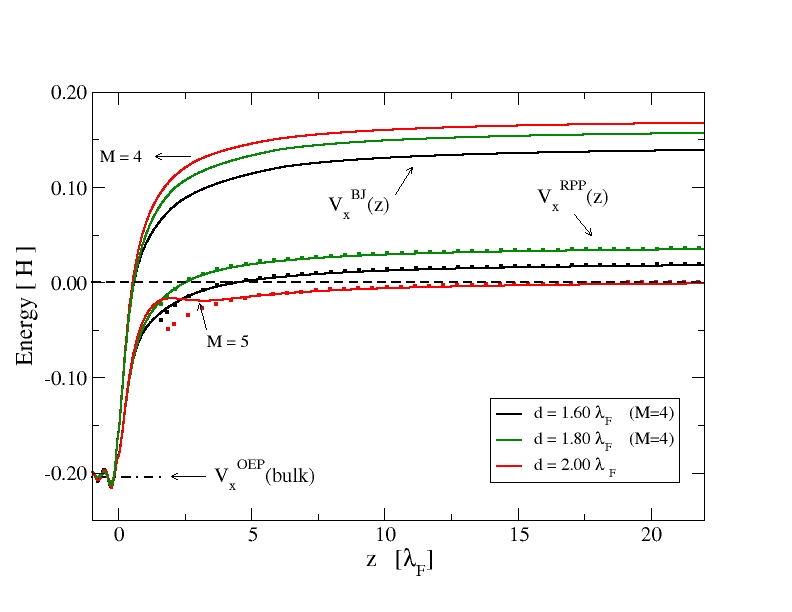}
 \caption{RPP and BJ exchange potentials for slab widths
 $d = 1.60, 1.80, \; \text{and} \; 2.00 \; \lambda_F$. Dotted lines corresponds to the RPP asymptotic expression in Eq.~(\ref{BJ-RPP-inf}). $r_s=3$, and $M = 4$ except for the slab width  $d = 2 \; \lambda_F$ in the RPP approximation that has  $M = 5$. In all cases, the metal-vacuum interface is at $z=0$. }
 \label{Fig:BJ-RPP}
\end{figure}
 
\begin{figure}  
 \includegraphics[width=12cm]{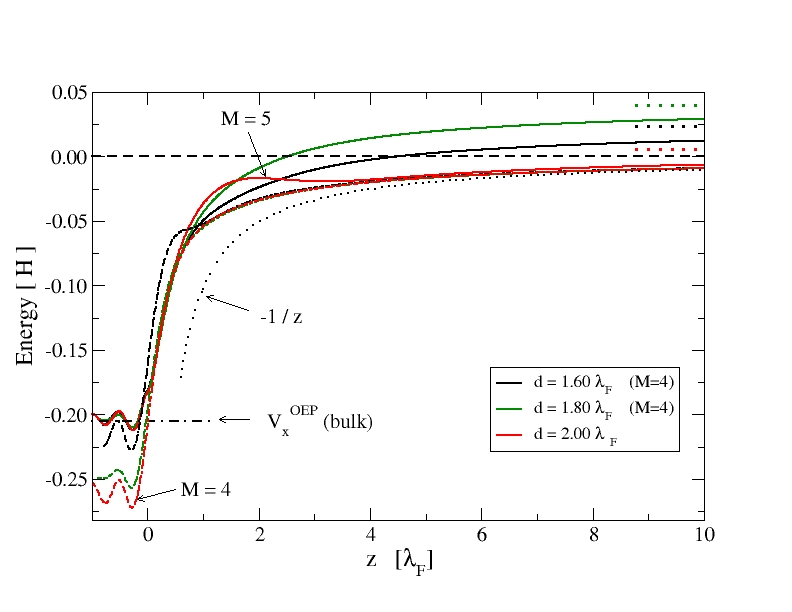}
 \caption{RPP (full lines) and $x$-only OEP exchange (dashed lines) potentials, for different slab widths $d$. 
 The dotted line corresponds to the universal limit $-1/z$. The upper right dotted segments are the $V_{\infty}^{\text{RPP}}$ corresponding limits.
 $r_s = 3$, and in all cases, the metal-vacuum interface is at $z=0$.}
 \label{Fig:RPP-OEP}
\end{figure}

In the context of the present jellium-slab system, the RPP model potential reads:
\begin{equation}
 V_{x,\sigma}^{\text{RPP}}(z) = V_{x,\sigma}^{\text{BR}}(z) + 
 C \left[ \frac{D_{\sigma}(z)}{n_{\sigma}(z)} \right ]^{1/2} \; .
 \label{BJ-RPP}
\end{equation}

The two main advantages of this model potential, as compared to its BJ counterpart, are the following: (i) it reproduces the correct asymptotic limit $-1/r$ for {\it any} finite $N$-electron system like atoms and molecules, and (ii) it is exact for {\it all} one-particle systems and not only for the hydrogen atom. Besides, if $D_{\sigma}(\mathbf{r})$ is taken as suggested in 
Eq.~(7) of Ref.~[\onlinecite{RPP10}] instead of Eq.~(\ref{Ds}) above, the exchange potential becomes gauge-invariant, which is particularly valuable when finite
systems are subject to electric or magnetic fields.
This model potential has been successfully tested for a variety of finite systems, including atoms, molecules, and atomic chains.\cite{ORPM10}
More recently, the RPP exchange functional has shown very promissory outputs, when compared with other several semilocal functionals, in a
large-scale DFT study on the influence of the exchange-correlation functional in the calculation of electronic band gaps of solids.\cite{Borlido20}
A version of the RPP model potential valid for low-dimensional systems has also
been suggested, successfully tested, and
proven to be very accurate in comparison with the corresponding OEP low-dimensional exchange potential.\cite{PRP10}

Our main interest here is to see how $V_{x,\sigma}^{\text{RPP}}(z)$ differs from $V_{x,\sigma}^{\text{BJ}}(z)$ in the vacuum asymptotic limit, for our present jellium-slab model of a metal surface.
From Eq.~(\ref{Dsa}), one obtains 
\begin{equation}
 V_{x,\sigma}^{\text{RPP}}(z/d \gg 1) \rightarrow V_{x,\sigma}^{\text{BR}}(z/d \gg 1)
 + \frac{C}{\sqrt{2}}  k_F^{M_{\sigma}} = -\frac{\alpha^{\text{BR}}}{z} + V_{\infty}^{\text{RPP}}\; ,
 \label{BJ-RPP-inf}
\end{equation}
with $V_{\infty}^{\text{RPP}} = C k_F^{M_{\sigma}} / \sqrt{2}$. 
While for finite systems the RPP exchange potential goes asymptotically towards the correct 
limit $-1/r$, in the case of a jellium slab we obtain a finite correction term, reflecting once 
more the hybrid  finite/extended spatial character of the slab geometry.
Asymptotically
($z/d \gg 1$), the BJ and RPP model potentials both tend to a positive material-dependent constant, which in the case of the RPP potential is proportional to $k_F^M$. In the particular case of a slab width corresponding to the HOSDL being just occupied ($\eta_M\sim 0^+$; $k_F^M\to 0$), $V_{x}^{\text{BR}}(z)$ and $V_{x}^{\text{RPP}}(z)$
both yield the correct $-1/z$ slab asymptotics, while $V_{x}^{\text{BJ}}(z)$ still yields a positive constant far into the vacuum.


The RPP model potential $V_{x}^{\text{RPP}}(z)$ is displayed in
Figs.~\ref{Fig:BJ-RPP} and \ref{Fig:RPP-OEP}, together with
$V_{x}^{\text{BJ}}(z)$ (in Fig.~\ref{Fig:BJ-RPP}) and $V_{x}^{\text{OEP}}(z)$
(in Fig.~\ref{Fig:RPP-OEP}). Both the BJ and RPP models reproduce, by construction, the correct bulk limit. On the other hand, they both fail to describe the actual exchange potential on the vacuum side of the surface, although the deviation is not so large in the case of the RPP model potential, and the work functions are, therefore, closer (although still too large) to their OEP counterparts (quoted above):
$W^{\text{RPP}}(d=1.6 \; \lambda_F) \simeq 0.1345$,
$W^{\text{RPP}}(d=1.8 \; \lambda_F) \simeq 0.1365$, and $W^{\text{RPP}}(d=2.00 \; \lambda_F) \simeq 0.1286$.

A shoulder in $V_{x}^{\text{RPP}}(z)$ is visible for $d=2 \;
\lambda_F$. The reason for this is that for this particular slab width
the fifth SDL is just occupied ($M=5$ and $\eta_5$ small), so that $k_F^5$ is small
($k_F^5 \sim 0.042$) and, therefore, large distances are needed to reach
$k_F z \gg 1$ and the correct asymptotics, as given by Eq.~(\ref{BJ-RPP-inf}). This is comparable to the
shoulder exhibited (also for $d=2 \; \lambda_F$) by $V_x^{\text{OEP}}(z)$, which is
visible in Fig. 2 for small values of $\eta_6$.

\begin{figure}  
 \includegraphics[width=12cm]{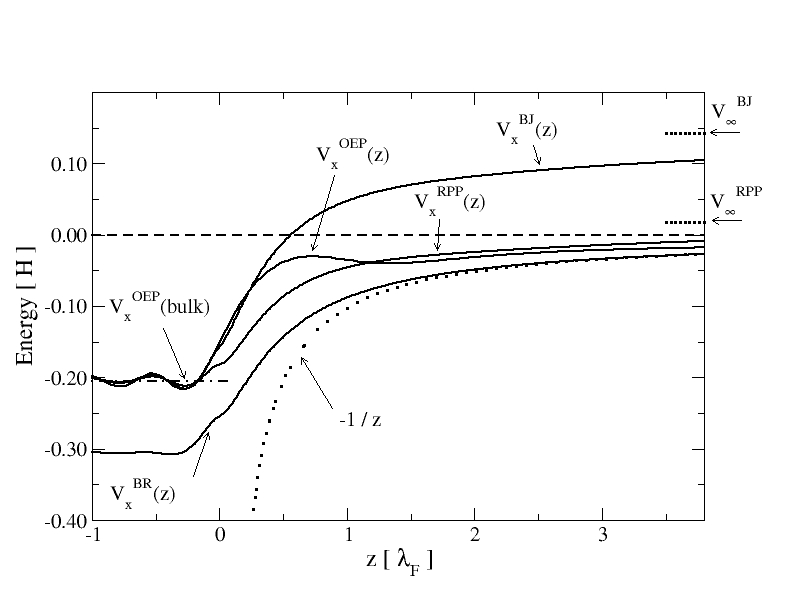}
 \caption{Self-consistent $V_x^{\text{OEP}}(z)$, $V_x^{\text{BR}}(z)$, $V_x^{\text{BJ}}(z)$ and $V_x^{\text{RPP}}(z)$ potentials. The dotted line corresponds to the universal limit $-1/z$.
 The upper right dotted segments are $V_{\infty}^{\text{RPP}} = 0.018$ and $V_{\infty}^{\text{BJ}} = 0.142$.
 $r_s=3$, $d=2.5785 \; \lambda_F$. In all cases, $M=6$, and the metal-vacuum interface is at $z=0$.}
 \label{Fig:all}
\end{figure}

As a sort of preliminary graphical conclusion, we display in Fig.~\ref{Fig:all} the three model potentials under study, together with their OEP counterpart, for a slab with $d=2.5785 \; \lambda_F$.
For this particular jellium slab, $V_x^{\text{OEP}}(z)$ somehow interpolates between $V_x^{\text{BJ}}(z)$ 
and $V_x^{\text{RPP}}(z)$ in the bulk (and in the neighborhood of the interface) and $V_x^{\text{RPP}}(z)$ far into the vacuum. $V_x^{\text{BJ}}(z)$ cleary fails to describe $V_x^{\text{OEP}}(z)$, except in the 
bulk region. Concerning $V_x^{\text{BR}}(z)$, it fails badly in the bulk, but it approaches the 
correct $-1/z$ asymptotics with considerable accuracy ($\alpha^{\text{BR}}$ being, in most cases, 
quite close to unity).

\section{Conclusions}
We have carried out a numerical and analytical study of the asymptotic behavior, the satisfaction of exact constraints, finite-size effects, and the work function of three semilocal approximations to the KS exchange potential of DFT, as applied to the jellium-slab model of a metal-vacuum interface, and we have analyzed the performance of these three model potentials by taking the exchange OEP as a reference.

In the case of the Becke-Roussel model, we have found $V_x^{\text{BR}}(z/d \gg 1) \rightarrow -\; \alpha^{\text{BR}}/z$, with $\alpha^{\text{BR}} < 1$, 
but in most cases close to unity. Regarding the wide-spread Becke-Johnson model potential, we have found that 
$V_x^{\text{BJ}}(z/d \gg 1) \rightarrow -\; \alpha^{\text{BR}}/z + V_{\infty}^{\text{BJ}}$,
with $V_{\infty}^{\text{BJ}} > 0$, which leads to a considerable overestimation of the work function, typically by a factor of two. Similar asymptotics are found for the RPP model potential: $V_x^{\text{RPP}}(z/d \gg 1) \rightarrow -\; \alpha^{\text{BR}}/z + V_{\infty}^{\text{RPP}}$,
but now with $V_{\infty}^{\text{RPP}}$ (also positive) being considerably smaller
than in the case of the BJ model potential. As a result, the RPP model potential is asymptotically closer (than its BJ counterpart) to the actual (OEP) KS exchange potential. Some finite-size features of the OEP are also exhibited by the RPP model potential, so we conclude that its performance, for jellium slabs, is superior to the performance of the other model potentials under study, and we suggest, therefore, its use for the {\it ab-initio} study of the electronic structure of real metal surfaces.
Besides, and considering the hybrid dimensionality of the slab geometry, which is finite along 
$z$ but extended in the perpendicular plane, we also suggest the use of the RPP semilocal 
exchange potential both for bulk and finite systems alike.

A natural follow-up of this work will be to explore the properties of these semilocal exchange potentials for the semi-infinite 
geometry, considering that in this case, due to its continuous energy spectrum, 
the asymptotic collapse towards the highest occupied slab discrete level employed here is 
not valid anymore.
Work is in progress along this line of research.

This work should also serve as a basis to further include the correlation contribution to the 
surface asymptotics. This represents a delicate issue for energy functionals, due to LDA error 
cancellations,~\cite{pe01} which means that improvements of the exchange functional are not 
beneficial unless they are accompanied with improvements on the correlation functional at the 
same level of approximation. This is not the case, however, in general, for the exchange-correlation 
potential outside a metal surface, since neither the LDA exchange potential nor the LDA correlation 
potential contribute to the actual asymptotics.

\section{acknowledgements}
 We thank UnCaFiQT (SNCAD) for computational resources.
C.M.H. wishes to acknowledge the financial support received from CONICET of Argentina through PIP 2014-47029.
C.R.P. wishes to acknowledge the financial support received from CONICET and ANPCyT of Argentina through grants PIP 2014-47029 and PICT 2016-1087.

\begin{appendix}
\section{Derivation of asymptotic expressions for $x_{\sigma}(z)$ and $b_{\sigma}(z)$}
\subsection{The vacuum limit $x_{\sigma}(z/d \gg 1)$ and the bulk limit $x_{\sigma} (z \rightarrow -\infty)$}
\begin{figure}
 \includegraphics[width=12cm]{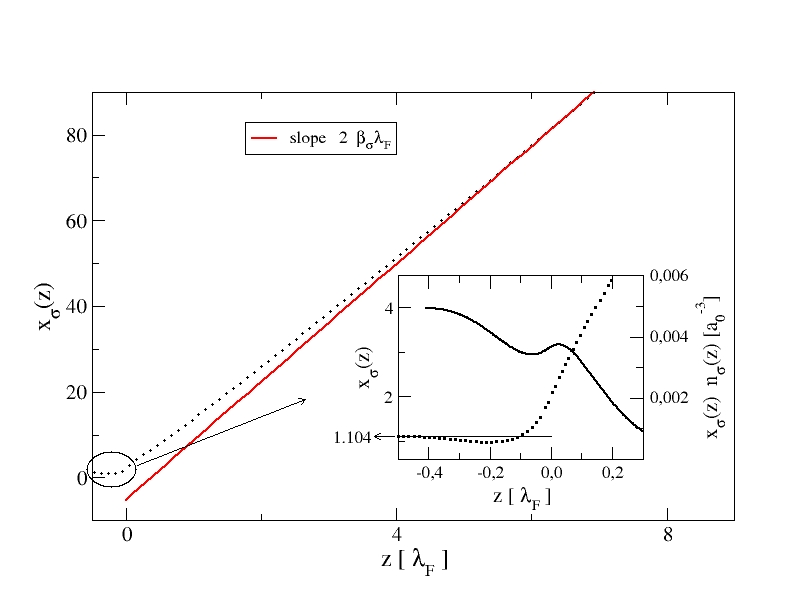}
 \caption{Numerical solution of Eq.~(A1) (dotted line) and its asymptotic limit given by Eq.~(A2) (red full line), for a slab with
 $r_s = 3$ and $d = 1.48 \; \lambda_F$.
 The inset shows the curve $x_{\sigma}(z)$ in the region close to the metal-vacuum interface (referred to the left vertical axis),
 and $x_{\sigma}(z) n_{\sigma}(z)$ (referred to the right vertical axis).}
 \label{Fig:x}
\end{figure}

We derive here the asymptotics of Eq.~(\ref{BR-3}) for a jellium-slab geometry. First of all, we rewrite Eq.~(\ref{BR-3}) as follows:
\begin{equation}
 x_{\sigma}(z) = - \; \frac{3}{2} \ln \left[ \frac{2\pi^{2/3}}{3}  \frac{n_{\sigma}^{5/3}(z)}{Q_{\sigma}(z)} 
 \frac{x_{\sigma}(z)-2}{x_{\sigma}(z)} \right] \; ,   
 \label{A1}
\end{equation}
and we then look for its asymptotic solution $x_{\sigma}(z \rightarrow \infty) =: x_{\sigma}(\infty)$.
Considering that $Q_{\sigma}(\infty) \sim n_{\sigma}(\infty)$, then
$n_{\sigma}^{5/3}(\infty) / Q_{\sigma}(\infty) \sim n_{\sigma}^{2/3}(\infty)$,
which goes exponentially to zero, making the argument inside the logarithm function arbitrarily small. 
Accordingly, $x_{\sigma}(\infty) \rightarrow \infty$, and the factor 
$[x_{\sigma}(\infty) - 2]/x_{\sigma}(\infty) \rightarrow 1$. By keeping the leading terms only, we find:
\begin{equation}
    x_{\sigma}(z \rightarrow \infty) \rightarrow -\;\frac{3}{2} \ln \left[n_{\sigma}^{2/3}(z \rightarrow \infty)\right] 
    \rightarrow - \;\frac{3}{2} \ln [\exp{(-4z\beta_{\sigma}/3)}] + B = 2 z \beta_{\sigma} + B \;,
    \label{A2}
\end{equation}
which proves that $x_{\sigma}(z)$ grows linearly with $z$ in the slab asymptotic region; $B < 0$ is a constant term dependent on the normalization parameter $A_i^{\sigma}$.
We show in Fig.~\ref{Fig:x} how the numerical solution of Eq.~(A1), which is valid for all values of $z$, coincides with the solution of Eq.~(A2) when $z$ is far outside into the vacuum.

For completeness, we now discuss the bulk solution ($z \rightarrow - \infty$, see Ref. [\onlinecite{Note2}]) of Eq.~(\ref{A1}), which leads to the determination of the parameter $\gamma$.
From Eq.~(\ref{BR-6}) and using 3D plane waves, one arrives at $t_{\sigma}({\text{bulk}}) = (3/5) (6\pi^2)^{2/3} n_{\sigma}(\text{bulk})^{5/3}$. From Eq.~(\ref{BR-5}), $D_{\sigma}({\text{bulk}})= t_{\sigma}({\text{bulk}})$, and introducing this into Eq.~(\ref{BR-4}), we find $Q_{\sigma}({\text{bulk}}) = - (\gamma/5) (6\pi^2)^{2/3} n_{\sigma}(\text{bulk})^{5/3}$.
Calling $x_{\sigma}(\text{bulk}) \equiv \bar{x}_{\sigma}$, the bulk version of Eq.~(\ref{BR-3}) reduces to
\begin{equation}
    \frac{\bar{x}_{\sigma}e^{-2\bar{x}_{\sigma}/3}}{\bar{x}_{\sigma}-2} = - \frac{10}{3(6\pi)^{2/3}} \frac{1}{\gamma} \; .
    \label{A3}
\end{equation}
This equation defines $\gamma$, once $\bar{x}_{\sigma}$ is known. For this, the physical constraint 
$V_{x,\sigma}^{\text{BR}}(\text{bulk}) = V_{x,\sigma}^{\text{Slater}}(\text{bulk})$ is imposed. Introducing this constraint and using Eq.~(\ref{BR-1}), one finds after some cancellations:
\begin{equation}
    \frac{1}{\bar{x}_{\sigma}e^{-\bar{x}_{\sigma}/3}} \left( 1- e^{-\bar{x}_{\sigma}} - \frac{\bar{x}_{\sigma}}{2} e^{-\bar{x}_{\sigma}} \right) =
    - \frac{3}{2} \left( \frac{3}{4\pi^2} \right)^{1/3} \; .
    \label{A4}
\end{equation}
By solving this equation numerically for $\bar{x}_{\sigma}$, we find that $\bar{x}_{\sigma} \simeq 1.104$; this is represented in Fig.~\ref{Fig:x} by a horizontal arrow on the vertical axis of the inset. From Eq.~(\ref{A3}), we find $\gamma \simeq 0.8$.\cite{BR89}
Interestingly, the determination of the parameter $\gamma$ is independent of the particular value 
of $r_s$, i.e., the electron density.

An analytical representation of the function $x_{\sigma}(\mathbf{r})$ that is valid for all distances can be found in Ref.~[\onlinecite{PGK08}]. 
In the present work, we have solved Eq.~(\ref{A1}) directly in a numerical way, without resorting 
to any analytical approximation.

\subsection{The limit $b_{\sigma}(z/d \gg 1)$}
We analyze here the asymptotic solution of Eq.~(\ref{BR-2}), which for a jellium-slab geometry can be written as follows:
\begin{eqnarray}
 b_{\sigma}(z) &=& \frac{{x_{\sigma}}(z) }{[8 \pi n_{\sigma}(z)]^{1/3}} e^{-{x_{\sigma}}(z)/3} \nonumber  \; , \\
               &=& \frac{1}{2} \left( \frac{2}{3} \right)^{1/2} \frac{n_{\sigma}^{1/2}(z)}{Q_{\sigma}^{1/2}(z)} x_{\sigma}(z) \; .
 \label{BR-2a}
\end{eqnarray}
Equation~(\ref{A1}) has been used in order to pass from the first to the second line of
Eq.~({\ref{BR-2a}}). We already have asymptotic expansions of
$n_{\sigma}(z)$ and $x_{\sigma}(z)$; but we still need to obtain the asymptotic expansion of $Q_{\sigma}(z)$. 
The first step for achieving this goal is to evaluate Eqs.~(\ref{ts})-(\ref{Qs}) in the limit $z/d \gg 1$. By restricting the sums over the SDL index $i$ to the HOSDL, i.e., to $i=M_{\sigma}$ for each spin component, one easily finds:
\begin{equation}
    t_{\sigma}(z/d \gg 1) \rightarrow \left[ \beta_{M_{\sigma}}^2 + \frac{(k_F^{M_{\sigma}})^2}{2} \right] n_{\sigma}(z/d \gg 1) \; ,
    \label{A6}
\end{equation}

\begin{equation}
    D_{\sigma}(z/d \gg 1) \rightarrow \frac{\left(k_F^{M_{\sigma}}\right)^2}{2} n_{\sigma}(z/d \gg 1) \; ,
    \label{Dsa}
\end{equation}
and
\begin{equation}
    Q_{\sigma}(z/d \gg 1) \rightarrow 
    \frac{1}{6} \left[ \frac{{\partial}^2}{{\partial z}^2} - \gamma \left(k_F^{M_{\sigma}}\right)^2 \right]  n_{\sigma}(z/d \gg 1) \; .
    \label{Qsa}
\end{equation}
Inserting the asymptotic density into Eq.~(\ref{Qsa}), we obtain:
\begin{equation}
    Q_{\sigma}^{1/2}(z/d \gg 1) \rightarrow \left( \frac{2}{3} \right)^{1/2} \beta_{M_{\sigma}}
    \left[1 - \frac{\gamma (k_F^{M_{\sigma}})^2}{4 \beta_{M_{\sigma}}^2}   \right]^{1/2} n_{\sigma}^{1/2}(z/d \gg 1) \; .
\end{equation}

Now we have everything we need for the evaluation of Eq.~(\ref{BR-2a}) in the asymptotic regime. We find:
\begin{equation}
    b_{\sigma}(z/d \gg 1) \rightarrow {z}\;{\left[1 - \frac{\gamma (k_F^{M_{\sigma}})^2}{4 \beta_{M_{\sigma}}^2}   \right]^{-1/2}} 
    \left(1 - \frac{1}{z \beta_{M_{\sigma}}} \right)\; .
    \label{bsa}
\end{equation}
It should be noted that Eq. (\ref{bsa}) includes the leading and the next-leading 
contributions to $b_{\sigma}(z/d \gg 1)$, with the later corresponding to the last factor. 
This next-leading contribution comes from not approximating the ratio 
$[x_{\sigma}(z/d \gg 1)-2] / x_{\sigma}(z/d \gg 1)$ by unity.

\end{appendix}


\begin{thebibliography}{[Vo]}
\bibitem{note10} The computational cost of a full OEP calculation is similar to that of hybrid DFT 
calculations.
\bibitem{note9} For very specific jellium slabs, when a new SDL has just been occupied ($\eta_M\sim 0; k_F^M\to 0$), the BJ model potential can be a reasonable approximation to the more realistic OEP (also positive in this case on the vacuum side of the surface); but for the great majority of slab widths the BJ model potential fails to describe the actual asymptotics, as in Fig.~\ref{Fig:BJ.OEP}.
\end{thebibliography}
\end{document}